\documentclass[a4paper,fleqn,usenatbib]{mnras}

\usepackage[T1]{fontenc}
\usepackage{ae,aecompl}

\usepackage{graphicx}	% Including figure files
\usepackage{amsmath}	% Advanced maths commands
\usepackage{amssymb}	% Extra maths symbols

\title[An Abel-inversion method for noisy data]{An efficient and flexible Abel-inversion method for noisy data}

\author[I. I. Antokhin]{
Igor I. Antokhin,$^{1}$\thanks{E-mail: igor@sai.msu.ru}
\\
$^{1}$Sternberg Astronomical Institute, M. V. Lomonosov Moscow State University, Moscow, 119992, Russian Federation\\
}

\date{Accepted XXX. Received YYY; in original form ZZZ}

\pubyear{2016}

\begin{document}
\label{firstpage}
\pagerange{\pageref{firstpage}--\pageref{lastpage}}
\maketitle

\begin{abstract}
We propose an efficient and flexible method for solving Abel integral equation of the first kind, frequently appearing in many fields of astrophysics, physics, chemistry, and applied sciences. This equation represents an ill-posed problem, thus solving it requires some kind of regularization. Our method is based on solving the equation on a so-called compact set of functions and/or using Tikhonov's regularization. A  priori constraints on the unknown function, defining a compact set, are very loose and can be set using simple physical considerations. Tikhonov's regularization on itself does not require any explicit a priori constraints on the unknown function and can be used independently of such constraints or in combination with them. Various target degrees of smoothness of the unknown function may be set, as required by the problem at hand. The advantage of the method, apart from its flexibility, is that it gives uniform convergence of the approximate solution to the exact solution, as the errors of input data tend to zero. The method is illustrated on several simulated models with known solutions. An example of astrophysical application of the method is also given.
\end{abstract}

\begin{keywords}
methods: numerical -- galaxies: kinematics and dynamics -- galaxies: individual: NGC~708, NGC~1129, NGC~1550
\end{keywords}

%%%%%%%%%%%%%%%%%%%%%%%%%%%%%%%%%%%%%%%%%%%%%%%%%%

%%%%%%%%%%%%%%%%% BODY OF PAPER %%%%%%%%%%%%%%%%%%

\section{Introduction}

\begin{figure}
\centering
\includegraphics[width=\columnwidth]{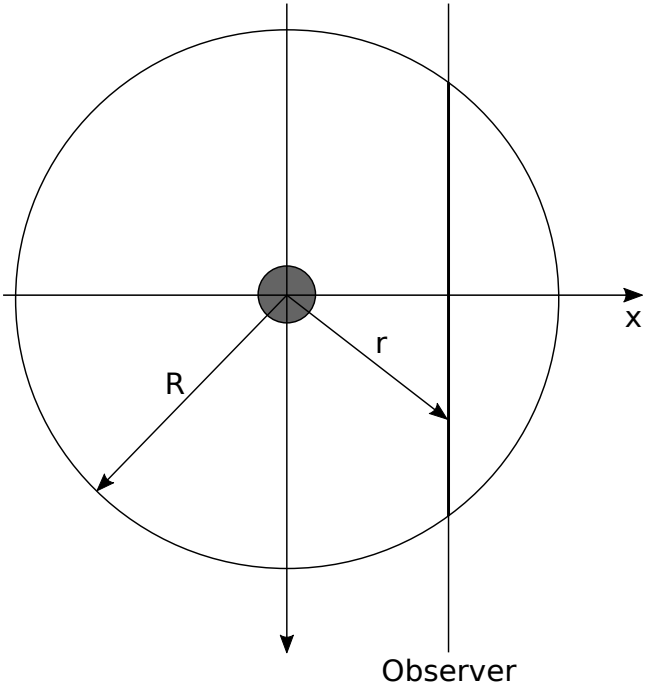}
\caption{Typical geometry of Abel equation in astrophysics. Note that in this particular geometry the right-hand part of \eqref{eq1} must be multiplied by 2.}
\label{fig1}
\end{figure}

Abel integral equation of the first kind first appeared in Abel's work dealing with the so-called `Abel mechanical problem', a generalization of the `tautochrone problem'. It frequently appears in many fields of astrophysics, physics, chemistry and applied sciences. In a slightly modified but equivalent form, it can be written as:

\begin{equation}
 \int_x^R \frac{z(r)r}{\sqrt{r^2-x^2}}\mathrm{d}r = u(x),
 \label{eq1}
\end{equation}

\noindent where $u(x)$ is the known input data and $z(r)$ is the unknown function. Equation \eqref{eq1} appears, e.g., whenever one measures an integral of some spherically or axially symmetric spatial function along a path (for instance, the line of sight), see Fig.\ref{fig1}. The spatial function may be an emission or absorption coefficient etc., and the measured function is correspondingly a surface brightness/absorption etc.

\begin{figure}
\centering
\includegraphics[width=\columnwidth]{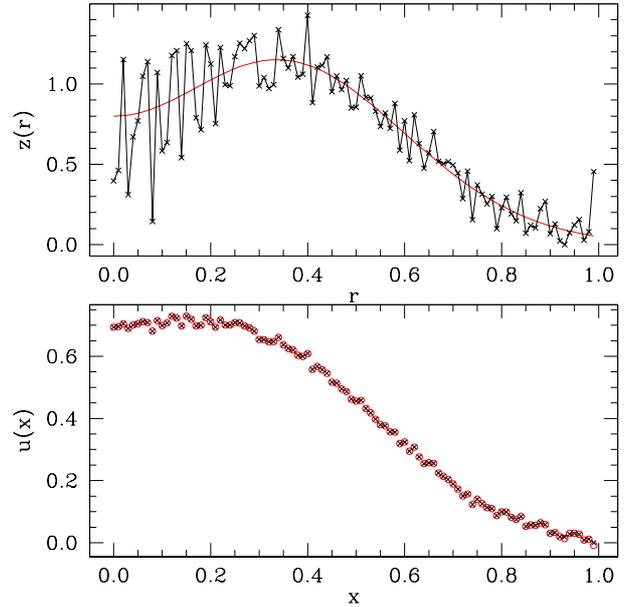}
\caption{An example of formal inversion of Abel equation. Exact and model $z(r)$ are shown by red solid line and black line/crosses respectively. Input and model $u(x)$ are shown by red circles and black crosses respectively.}
\label{fig2}
\end{figure}

Equation \eqref{eq1} allows a formal inversion to obtain $z(r)$. However, it includes the first derivative of $u(x)$ and is unusable in all practical cases when $u(x)$ contains noise. The equation presents an ill-posed problem. It can be shown \citep[see, e.g.,][] {craig_1979} that, in the presence of even infinitely small noise in $u(x)$, the errors in $z(r)$ obtained by the formal inversion can be unbounded. An example of such formal `solution' is shown in Fig.\ref{fig2}. An analytic function $z(r)$ (shown in the upper plot by solid red line) was used to obtain $u(x)$ by evaluating the integral in \eqref{eq1}. Then Gaussian noise with $\sigma=0.01$ was added to $u(x)$ and the resulting function was used as input data to \eqref{eq1}. Clearly, the `solution' fits not only the regular part of $u(x)$ but also noise. For a different noise sample, the `solution' will be completely different.

Several methods of solving \eqref{eq1} were developed in the past decade or two. The most frequently used approaches are approximation of the unknown function by a sum of, for instance, Chebyshev or Bernstein polynomials, splines, Gaussians, etc. (depending on a particular problem) or smoothing the input data $u(x)$, again using some polynomials. For references, see \citet{craig_1979}, \citet{Knill1993}, \citet{bend1991}, \citet{dixit2011}, etc. The key point in this approach is to truncate the degree of the polynomial approximation, so  the high-frequency components are cut off thus stabilizing the solution. The main drawbacks are that (i)they do not guarantee convergence of the numeric solution to the true solution as the accuracy of the input data increases, and (ii)the choice of the polynomial degree is rather arbitrary.

Another approach is to consider \eqref{eq1} as a convolution trasform \citep[e.g.,][]{Sumner56} and to solve it using, for instance, Fourier deconvolution. In this case, all problems related to noise in the data remain. Some smoothing is still required, see, e.g., \cite{Prasenjit96} who used a functional form for the unknown functon.

In astrophysics, first attempts of solving \eqref{eq1} used a formal inversion and hence numeric differentiation of $u(x)$  \citep[e.g.,][]{Plummer11}. Later on, \cite{Wallenquist1933} suggested an elegant numeric method. Let us briefly remind the reader its key points. The object (a globular cluster) is assumed to be spherically symmetric. It may be subdivided into a series of concentric spherical shells around its centre. The space star density within a shell is assumed to be constant. The shells are projected on the sky plane as concentric rings. The measured data are numbers of stars in the rings. The space star density in the outermost shell is the number of stars in the outer ring divided by the part of the outer shell's volume projected on the outer ring. Two shells, the outermost one and the last but one outermost shell, are projected on the last but one ring. The number of stars belonging to the outermost shell is the already known space star density in this shell multiplied by the part of its volume projected on the last but one ring. Subtracting this number from the total number of stars in the last but one ring, we get the number of stars of the last but one shell projected on this ring. Dividing it by the corresponding volume (that of the part of the last but one shell projected on the last but one ring), we get the space star density in the last but one shell. Then we proceed similarly from the edge of the cluster to its centre.

It is worth noting that, while being brilliantly simple, Wallenquist's method has an inherent shortcoming. As described above, it does not involve any regularization, so the errors of the input data are directly translated into the solution. The resulting space density distribution is not necessarily a smooth function, nor is it guaranteed to be, e.g., convex (as is probably expected), or even monotonically decreasing. This is why in his original paper, before applying his method, Wallenquist approximated the measured surface densities with some smooth functions, thus essentially using one of the approaches outlined above.

In our paper, \citet{ant_2012} (Paper I), we proposed a method for solving Fredholm integral equation of the first kind on so-called compact sets of functions combined with Tikhonov's regularization. This method can be applied to any linear integral equation of the first kind. In the current paper, we expand it to the case of Abel equation.

In section 2, we present numeric representation of equation \eqref{eq1} and briefly describe the solution method. In section 3, the method is illustrated on various artificial input data with known solutions. The influence of errors in the input data on the solution is also demonstrated. In section 4, an astrophysical example of using the method is presented. We summarize the results in Section 5.

\section{Numeric representation and the solution method}

The solution method developed in Paper I is designed for Fredholm equation,

\begin{equation}
 \int\limits_a^b K(x,s)z(s)\,\mathrm{d}s = u(x),  \quad s \in [a,b], x \in [c,d] \, ,
 \label{eq2}
\end{equation}

\noindent where $K(x,s)$ is the known kernel of the integral. The easiest way to apply the algorithms from that paper to the current case is to re-write equation \eqref{eq1} as follows:

\begin{equation}
 \int\limits_{\rm 0}^R K(x,r)z(r)\,\mathrm{d}r = u(x),
 \label{eq3}
\end{equation}

\noindent where

\begin{equation}
 \left\{ \begin{array}{rcl}
 K(x,r) & = & \frac{r}{\sqrt{r^2-x^2}} \quad \mathrm{if} \quad r \ge x \, ,\\
 K(x,r) & = & 0 \quad \mathrm{if} \quad r < x \, .
         \end{array}
 \right.
 \label{eq4}
\end{equation}

A tempting straightforward numeric approximation of \eqref{eq3} looks as follows. Let us introduce an even grid $\{r\}_{j=1}^n$ on $[0,R]$ with the step $h_r$. The function $z(r)$ then becomes an $n$-dimensional vector. The input data $u(x)$ can be measured on even or uneven grid. Let $u(x)$ be measured at $m$ points $x_i$, $i=1,...,m$. Then, according to the standard rectangle integration formula, \eqref{eq3} could be represented as:

\begin{equation}
 \sum_{j=1}^n K(x_i,r_j)z_jh_r = u(x_i) \, ,\quad i=1,...,m \, .
 \label{eq5}
\end{equation}

However, there is a difficulty in evaluating the sum in this expression: the integral kernel in equation \eqref{eq1} has an (integrable) singularity. At $r=x$, the denominator becomes zero. To overcome this difficulty, we instead use a so-called generalized formula of left rectangles. In a standard rectangle formula for numeric integration, one assumes that the integrand is constant within an integration grid cell. In the generalized formula, we assume that only $z(r)$ is constant (and equal to the $z_j$ value at the left end of the cell $[j,j+1]$). Then, one can move $z_j$ outside the integral and compute it analytically:

\begin{equation}
 \int\limits_{r_j}^{r_{j+1}} \frac{z(r)r}{\sqrt{r^2-x^2_i}}\mathrm{d}r \approx z_j\left( \sqrt{r^2_{j+1}-x^2_i}-\sqrt{r^2_j-x^2_i} \right) \, .
 \label{eq6}
\end{equation}

With this approximation, the final numeric representation of \eqref{eq1} becomes:

\begin{equation}
 A\boldsymbol{z} \equiv \sum_{j=1}^n a_{ij}z_j = u(x_i) \, ,\quad i=1,...,m \, ,
 \label{eq7}
\end{equation}

\noindent where

\begin{equation}
 \left\{ \begin{array}{rcl}
 a_{ij} & = & \left( \sqrt{r^2_{j+1}-x^2_i}-\sqrt{r^2_j-x^2_i} \right) \, , \, r_j \ge x_i\\
 a_{ij} & = & 0 \, , \, r_j < x_i \, .
         \end{array}
 \right.
 \label{eq8}
\end{equation}

One further note is necessary here. To be valid, the approximation defined by \eqref{eq7}, \eqref{eq8} must use a grid on $r$ which includes all points $x_i$, $i=1,...,m$. In other words, for every $x_i$ there must exist an index $j$ such that $r_j=x_i$. Otherwise, the part of the integral on the interval from $x_i$ to the smallest $r_j>x_i$ would be lost.

A detailed description of the solution method for equation \eqref{eq2} is given in Paper I. Here we briefly remind the reader the main points of the method. Solving \eqref{eq2} with noisy $u(x)$ requires a regularization algorithm. Parametrization of $z(r)$ is an example of such an algorithm. However, parametrization is rather restrictive and in some cases is impossible (no adequate parametric model is known). Instead of parametrization, some a priori constraints on $z(r)$ may be used to construct a regularization algorithm. \citet{Tikhonov95} showed that a stable solution of equation \eqref{eq2} can be obtained with a priori constraints defining a so-called compact set of functions (any parametric model is an example of such a set). Solution on a compact set consists in minimizing the deviation of the model from the data (the norm):

\begin{equation}
 ||A\boldsymbol{z}-\boldsymbol{u}||^2 \rightarrow \mathrm{min} \, .
 \label{eq9}
\end{equation}

What are the minimal a priori constraints defining a compact set? \citet{Tikhonov95} showed that one possible assumption was that the unknown function was non-negative and monotonically non-increasing. Other, more strict, examples are non-negative convex/concave/concave-convex  monotonically non-increasing functions. These assumptions are non-parametric and loose enough.

The drawback is that they do not require the solution to be smooth, which is often a natural property of a physical function.  \citet{Tikhonov95} introduced a functional (named after him):

\begin{equation}
\Phi(\alpha,\boldsymbol{z}) = ||A\boldsymbol{z}-\boldsymbol{u}||^2+\alpha||\boldsymbol{z}||^2 \, .
\label{eq10}
\end{equation}

\noindent The basic idea of Tikhonov's regularization is very simple: instead of minimizing the residual to find an approximate solution, one has to minimize Tikhonov's functional. The second term in this expression is the so-called stabilizing term. It is constructed in such a way that it is small when $\boldsymbol{z}$ is smooth and large when $\boldsymbol{z}$ is fluctuating. Thus, minimizing $\Phi(\alpha,\boldsymbol{z})$, one can minimize the residual while suppressing instability in $\boldsymbol{z}$. The regularization parameter $\alpha$ controls the relative weight of the stabilizing term. The recipe for choosing $\alpha$ is given in \citet{Tikhonov95} and in Paper I. In short, $\alpha$ must be chosen so that the first norm in \eqref{eq10} is equal to the uncertainty of the input data. Let us stress that Tikhonov's regularization does not require any explicit a priori constraints on the unknown function (its being monotonic etc.) apart from its smoothness. It can be used as a stand alone method for obtaining a stable solution of equation \eqref{eq1}. However, if there exist some reasonable constraints on $z(r)$, these can be combined with Tikhonov's regularization.

Particular expressions for the norms depend on which metric spaces are used for $A\boldsymbol{z}$, $\boldsymbol{u}$ and $\boldsymbol{z}$. For instance, in the $L_2$ space (this is the space of real functions square-integrable on $[a,b]$, the proximity of functions is closeness in the mean square), the metric is:

\begin{equation}
 \rho(x,y)=\left[\int\limits_a^b (x(s)-y(s))^2\,{\mathrm d}s\right]^{0.5}, \nonumber
\end{equation}

\noindent so

\begin{equation}
 ||z||^2_{L_2} = \int\limits_0^R z^2(r)\,{\mathrm d}r \, . \nonumber
\end{equation}

In the $W_2^1$ space (the space of real functions, the functions themselves and their first derivatives are square-integrable on $[a,b]$), the metric is:

\begin{equation}
 \rho(x,y)=\left[\int\limits_a^b(x(s)-y(s))^2\,{\mathrm d}s+\int\limits_a^b(x'(s)-y'(s))^2\,{\mathrm d}s\right]^{0.5} \, , \nonumber
\end{equation}

\noindent so 

\begin{equation}
 ||z||^2_{W_2^1} = \int\limits_0^R[z^2(r)+z'(r)^2]\,{\mathrm d}r \, . \nonumber
\end{equation}

If necessary, even better smoothness can be achieved using the $W_2^2$ space ($||z||^2_{W_2^2} = \int_0^R[z^2(r)+z'(r)^2+z''(r)^2]\,{\mathrm d}r$). 

Equation \eqref{eq1} is a linear problem, so the functionals to minimize in \eqref{eq9}, \eqref{eq10} are quadratic. In this case, an effective minimization algorithm with a priori constraints on $z(r)$ is the conjugate gradient method, combined with a projection method such as that described in \citet{Rosen60}, \citet{Himmelblau72}.

Further details of the algorithm as well as numeric representations of various a priori constraints on $z(r)$ etc. are given in Paper I.

\begin{figure*}
\centering
\includegraphics[width=0.3\textwidth]{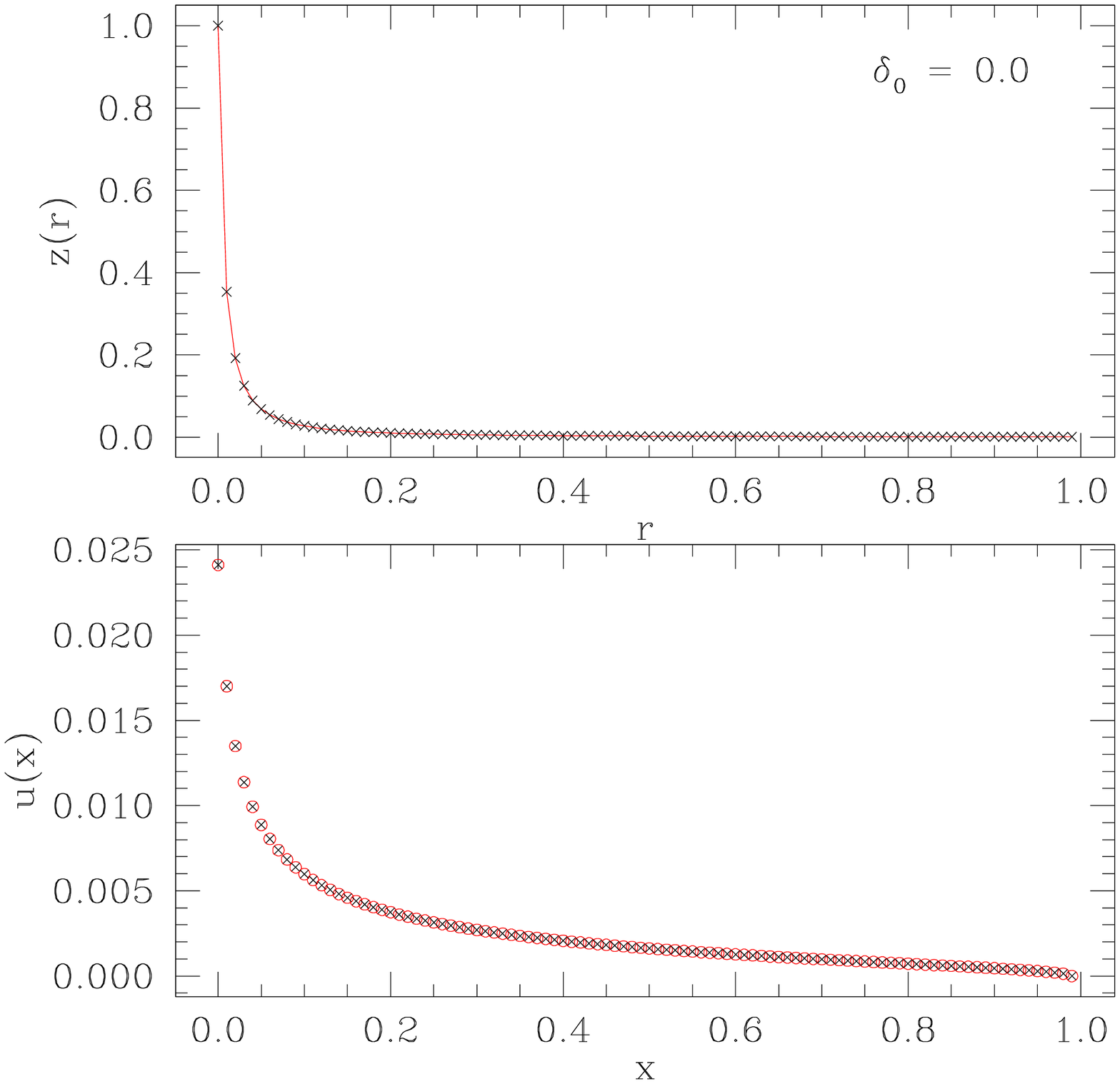}
\includegraphics[width=0.3\textwidth]{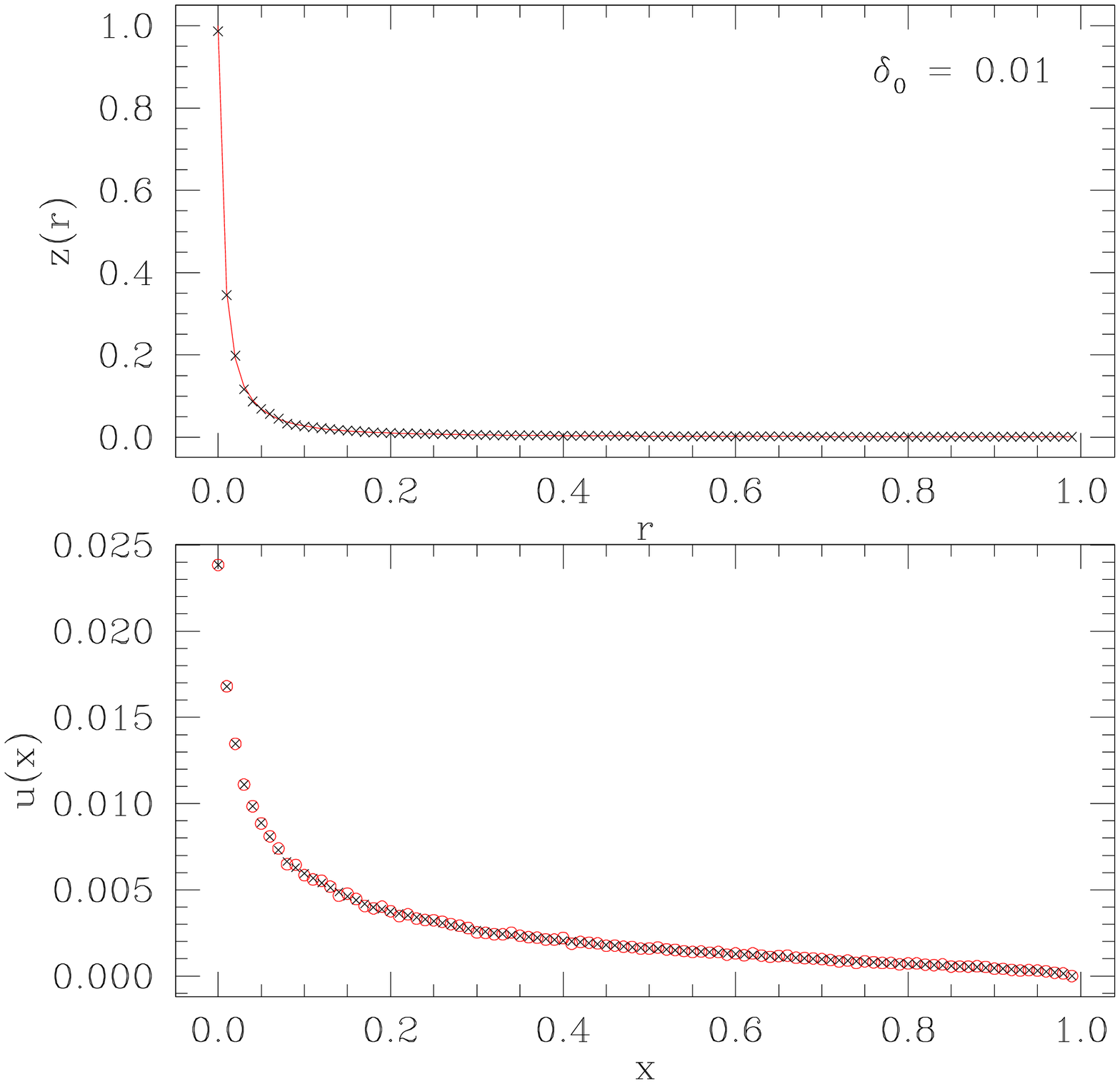}
\includegraphics[width=0.3\textwidth]{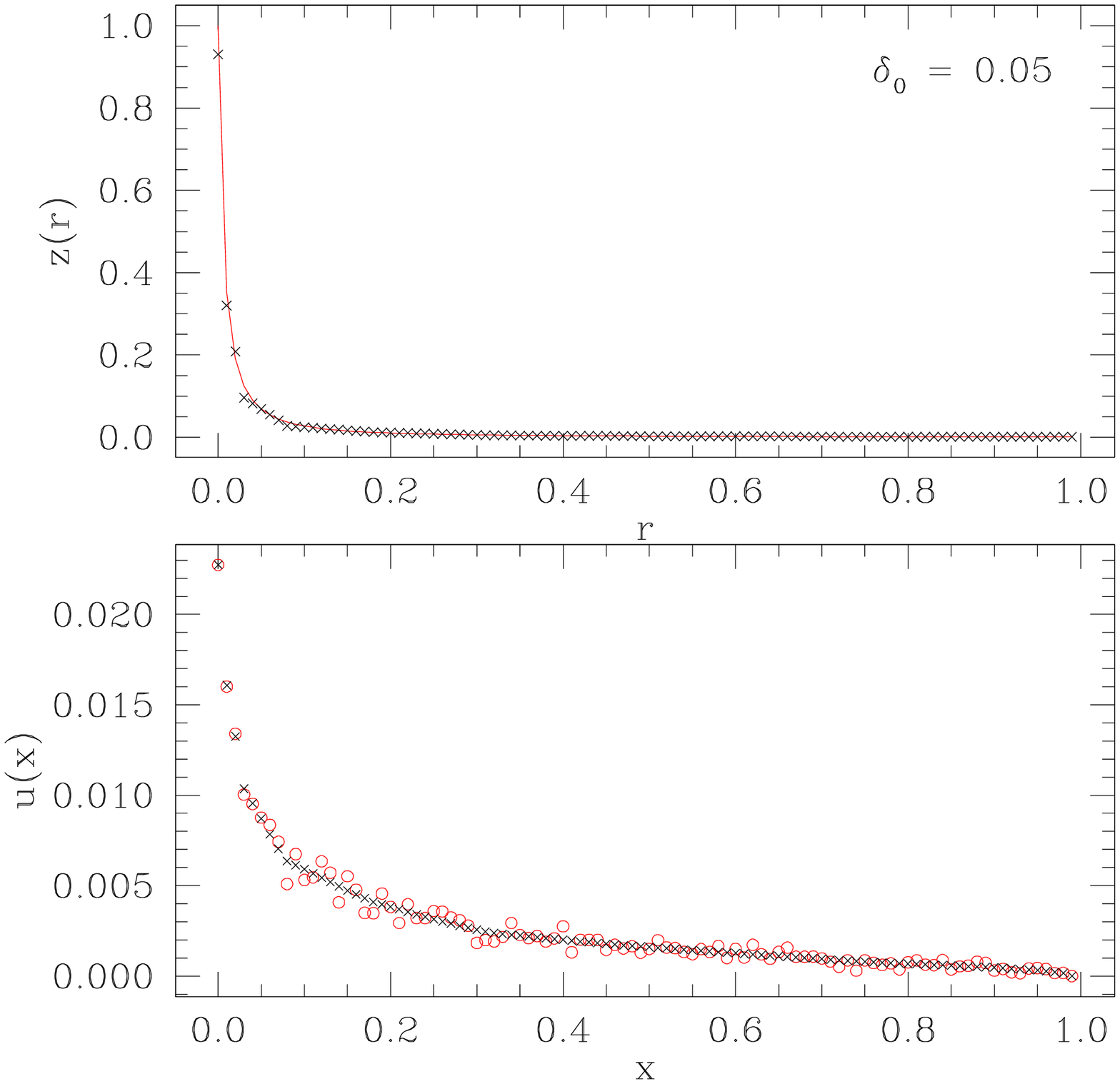}
\caption{Model 1 for various values of $\delta_0$ ($\delta_0=0.0$ means that exact $u(x)$ was used as input). In all plots, input $u(x)$ and exact $z(r)$ are plotted as red circles and red solid lines respectively, while the corresponding model functions are shown as black crosses.}
\label{fig3}
\end{figure*}

\begin{figure*}
\centering
\includegraphics[width=0.245\textwidth]{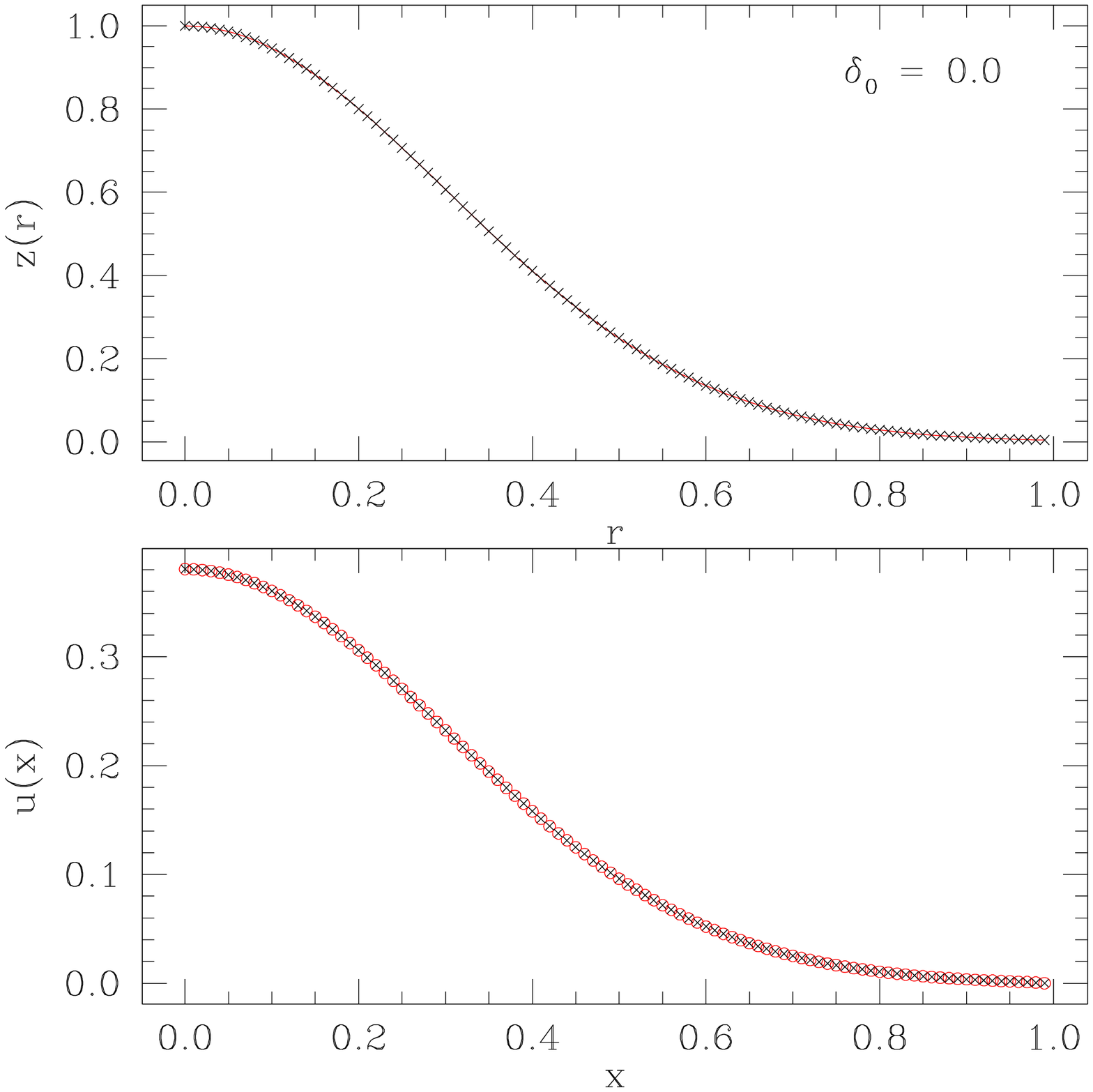}
\includegraphics[width=0.245\textwidth]{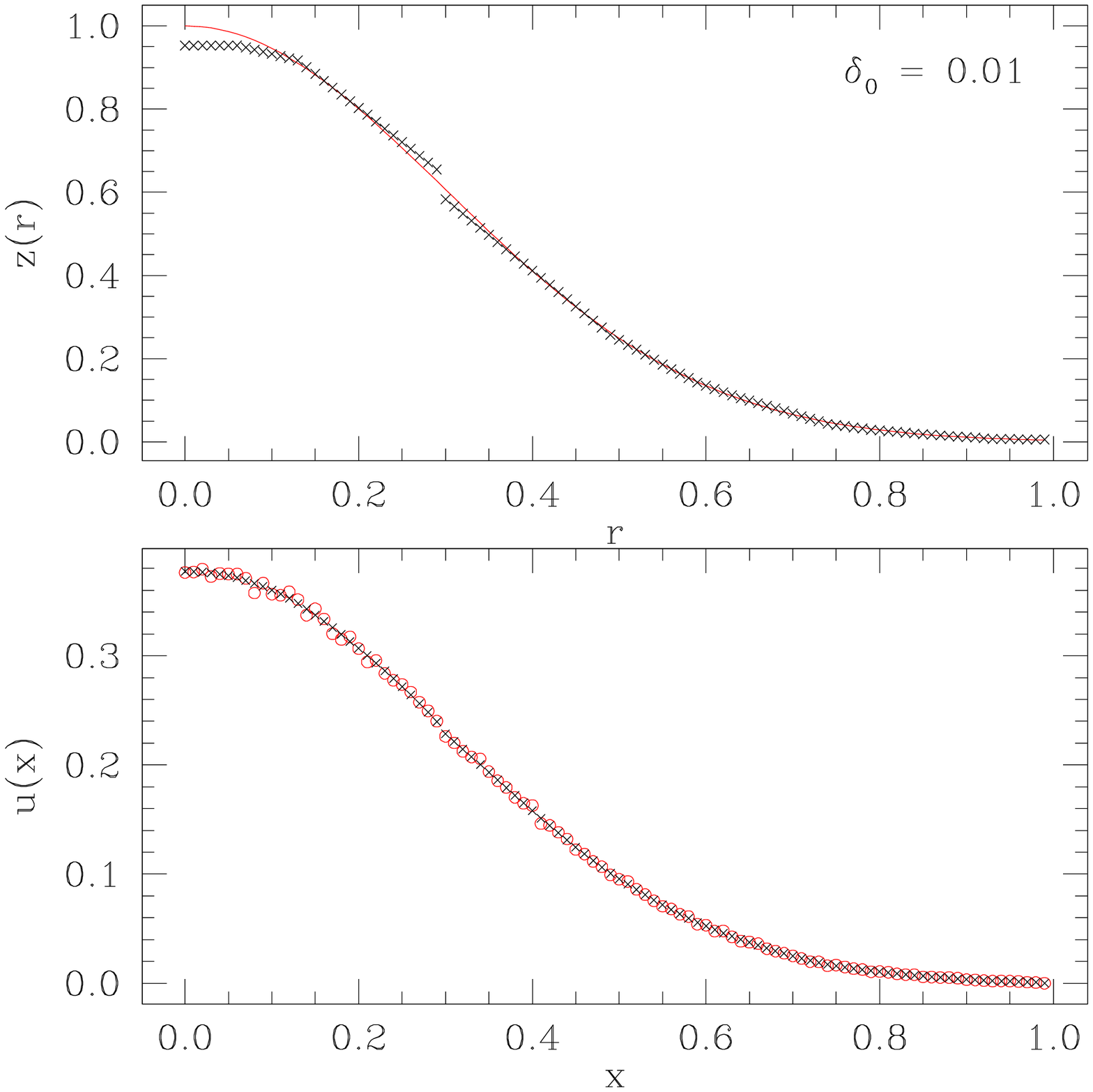}
\includegraphics[width=0.245\textwidth]{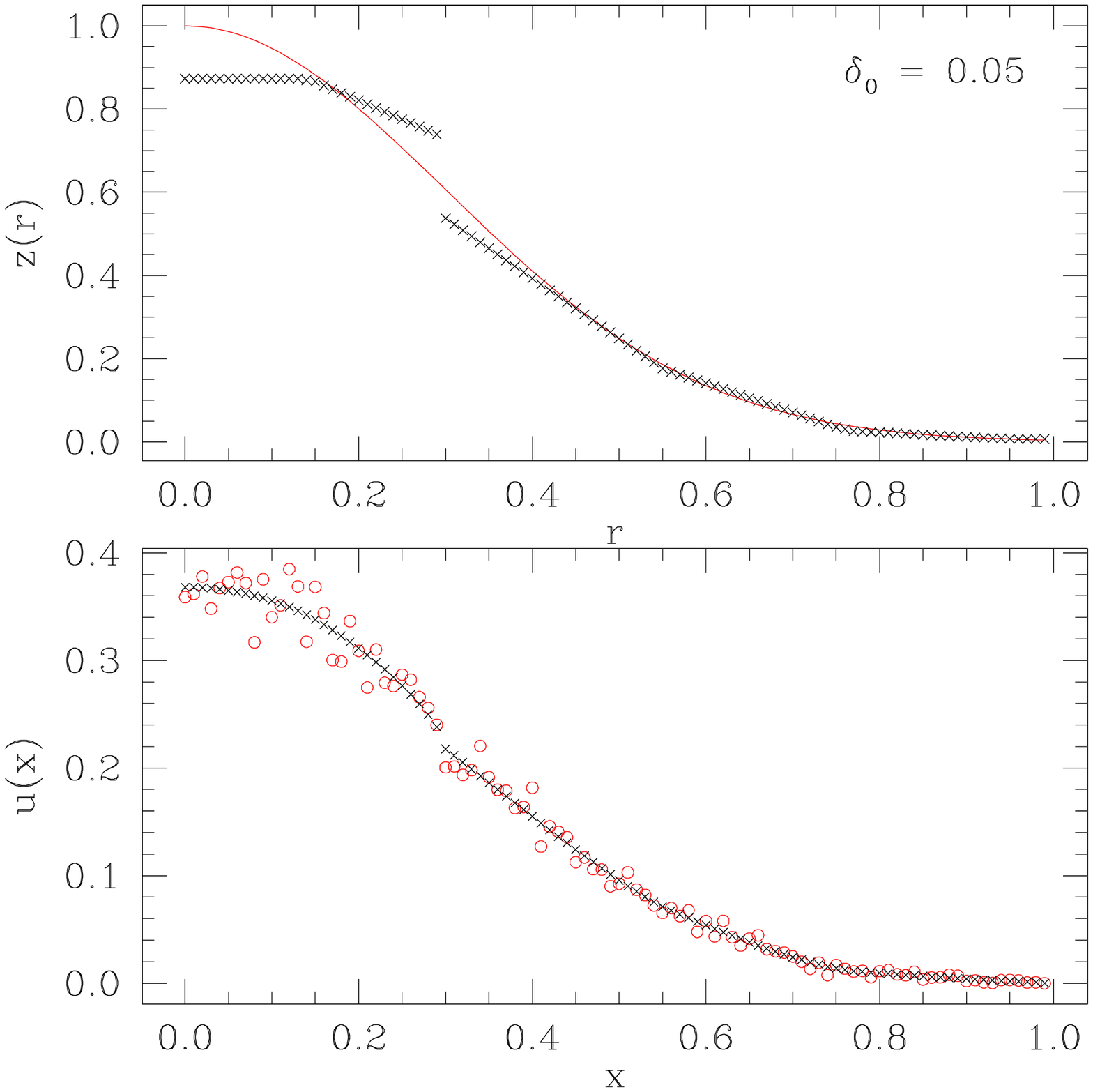}
\includegraphics[width=0.245\textwidth]{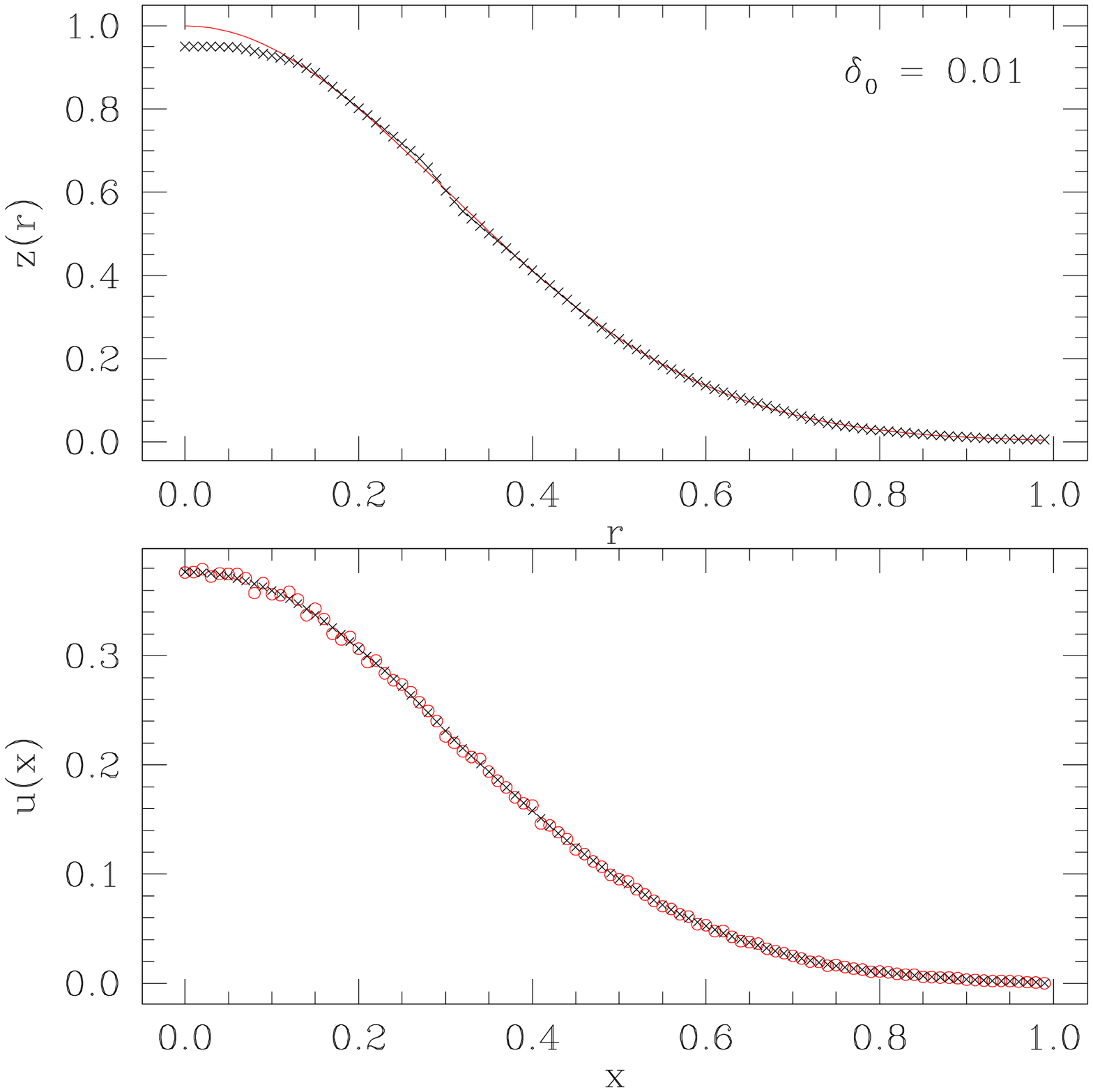}
\caption{Model 2, left to right. Three solutions on the compact set of concave-convex monotonically non-increasing functions without Tikhonov's regularization for $\delta_0$ equal to $0$ (exact $u(x)$ as input data), $0.01$,  and $0.05$. The last plot shows the solution for $\delta_0=0.01$ with added Tikhonov's regularization. The meaning of the colors and dot/line styles is the same as in Fig.\ref{fig3}.}
\label{fig4}
\end{figure*}

\section{Simulated problems}

We illustrate the method on three simulated problems with known solutions. In all cases, $u(x)$ was obtained by computing the integral in \eqref{eq1} with an analytic $z(r)$. Then, calculated $u(x)$ was used as input for the method. Even grids with the number of knots $n=m=100$ were used in numeric representation of $u(x)$ and $z(r)$. $R$ was set to $1.0$.

To show the influence of input errors on the solution, we added Gaussian noise to $u(x)$ in the following way. Very often (at least in astronomy), $u(x)$ is a measured flux. In this case, $u(x)$ is affected by Gaussian (strictly speaking, Poisson) noise proportional to $\sqrt{u(x)}$. As $u(x)$ in our simulations is an arbitrarily scaled function, let us normalize the noise distribution as follows. Let the relative noise of $u(0)$ be $\delta_0$. From Poisson statistics, we know that $\delta_0 = 1/\sqrt{N}$, where $N$ is the signal (e.g., the number of photons). In our case, one can write \mbox{$N=S\cdot u(0)$}, where $S$ is the required scaling factor. Thus, $S=1/(\delta_0^2u(0))$. To introduce some noise to exact $u(x_i)$, one has to draw a random number from the Gaussian distribution with the zero mean and standard deviation $\sigma(x)=\sqrt{S\cdot u(x_i)}$ and add this value to $u(x_i)$. In examples below, we show solutions for various values of $\delta_0$.

\subsection{Model 1: power law}

In this model, $z(r)$ is assumed to be a power law function:

$$
  z(r) = \frac{0.01^{3/2}}{(r+0.01)^{3/2}}\,\, , r \in [0,1] \, ,
$$

\noindent where the $0.01$ constant is added to avoid infinity at $r=0$. Power law functions can be expected in various astronomical objects. One example (albeit non directly observable) is quasi-spherical accretion onto a neutron star; the density of the accreting material is expected to drop with distance from the star as a power law with degree $3/2$ \citep{Shakura12}.

In Fig.\ref{fig3}, solutions of the problem are shown for various values of noise added to exact $u(x)$. Equation \eqref{eq1} was solved on the compact set of convex monotonically non-increasing functions without Tikhonov's regularization.

\subsection{Model 2: Gaussian function} 

In this model, $z(r)$ is a Gaussian function:

$$
z(r) = e^{-\frac{r^2}{2 \sigma^2}} \,\, , \sigma=0.3, \,\, , r \in [0,1] \, ,
$$

A Gaussian was chosen as an example of a concave-convex function. In astrophysics, it may represent, e.g., an object with a dense core and rarefied outer part.

In the first three plots in Fig.\ref{fig4}, solutions of the problem are shown for exact $u(x)$ and for $u(x)$ with added noise, $\delta_0=0.01$ and $0.05$. Equation \eqref{eq1} was solved on the compact set of concave-convex monotonically non-increasing functions without Tikhonov's regularization. The location of the inflection point was a free parameter. It is defined by the minimum of the deviation of a model solution from input data (the norm). The noise values, $0.01$ and $0.05$, are possibly an exaggeration. With modern instrumentation, one could expect that the relative accuracy of real data can be significantly better than 5 or 1 per cent. The purpose of these values is to show the qualitative behaviour of the numeric solution, when errors of input data change.

Clearly, when errors in $u(x)$ increase, the numeric solution deviates from the exact one more and more. It also becomes less smooth. To obtain a smoother solution, the problem was solved with the same a priori constraints as before, with added Tikhonov's regularization in the metric space $W_2^1$. To save journal space, only the case with  $\delta_0=0.01$ is shown in Fig.\ref{fig4} (the rightmost plot).

\subsection{Model 3: non-monotonic function} 

The purpose of this model is to demonstrate the method when the unknown function cannot be assumed to be monotonic, convex etc. In this case, only Tikhonov's regularization can be used. Exact $z(r)$ is defined as a product of a Gaussian function and a parabola:

$$
z(r) = 4 (3r^2+0.2) e^{-\frac{r^2}{2 \sigma^2}} \,\, , \sigma=0.3, \,\, r \in [0,1] \, .
$$

After integration, Gaussian noise with $\delta_0=0.01$ was added to $u(x)$. 

\begin{figure}
\centering
\includegraphics[width=0.49\columnwidth]{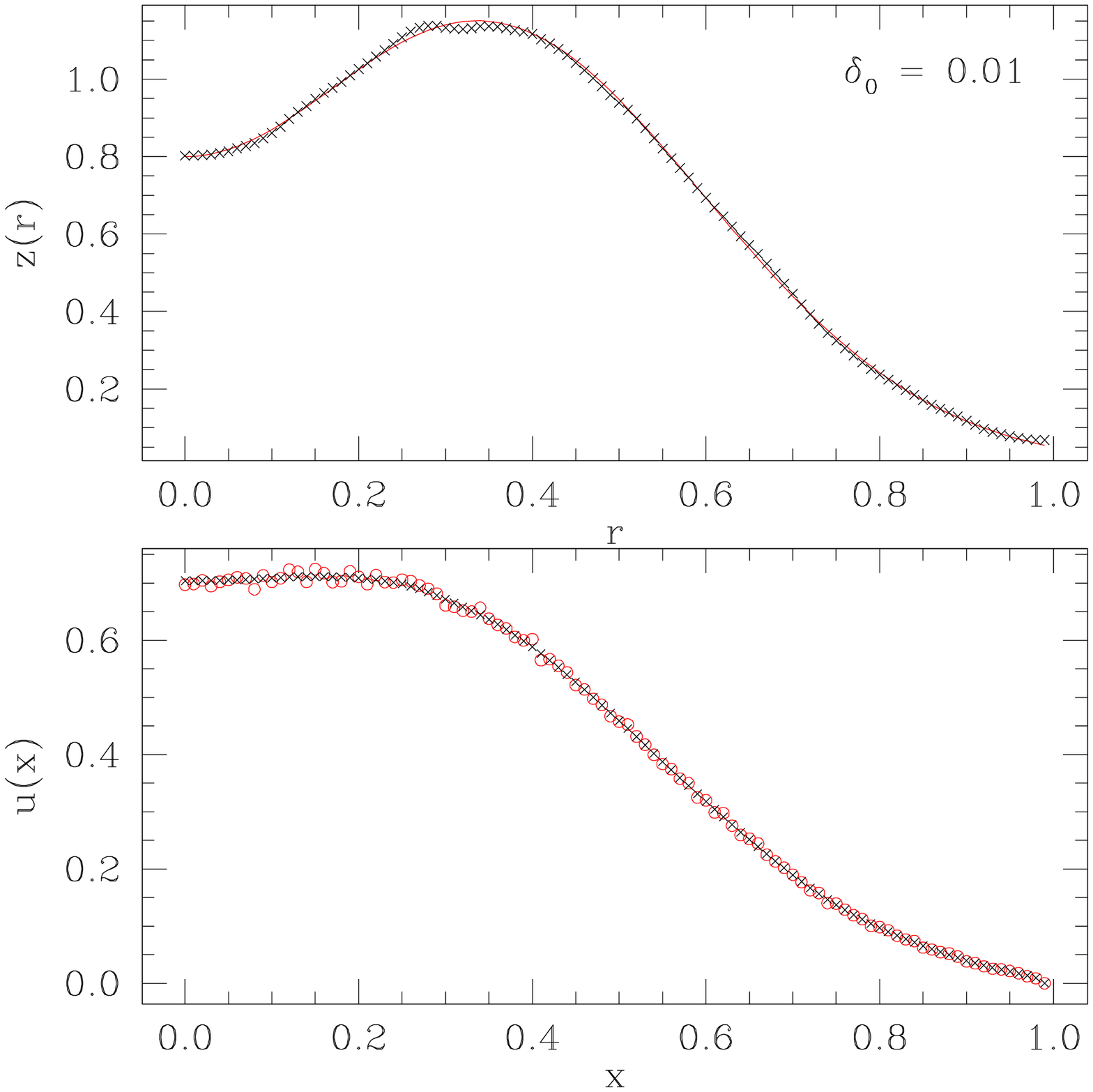}
\includegraphics[width=0.49\columnwidth]{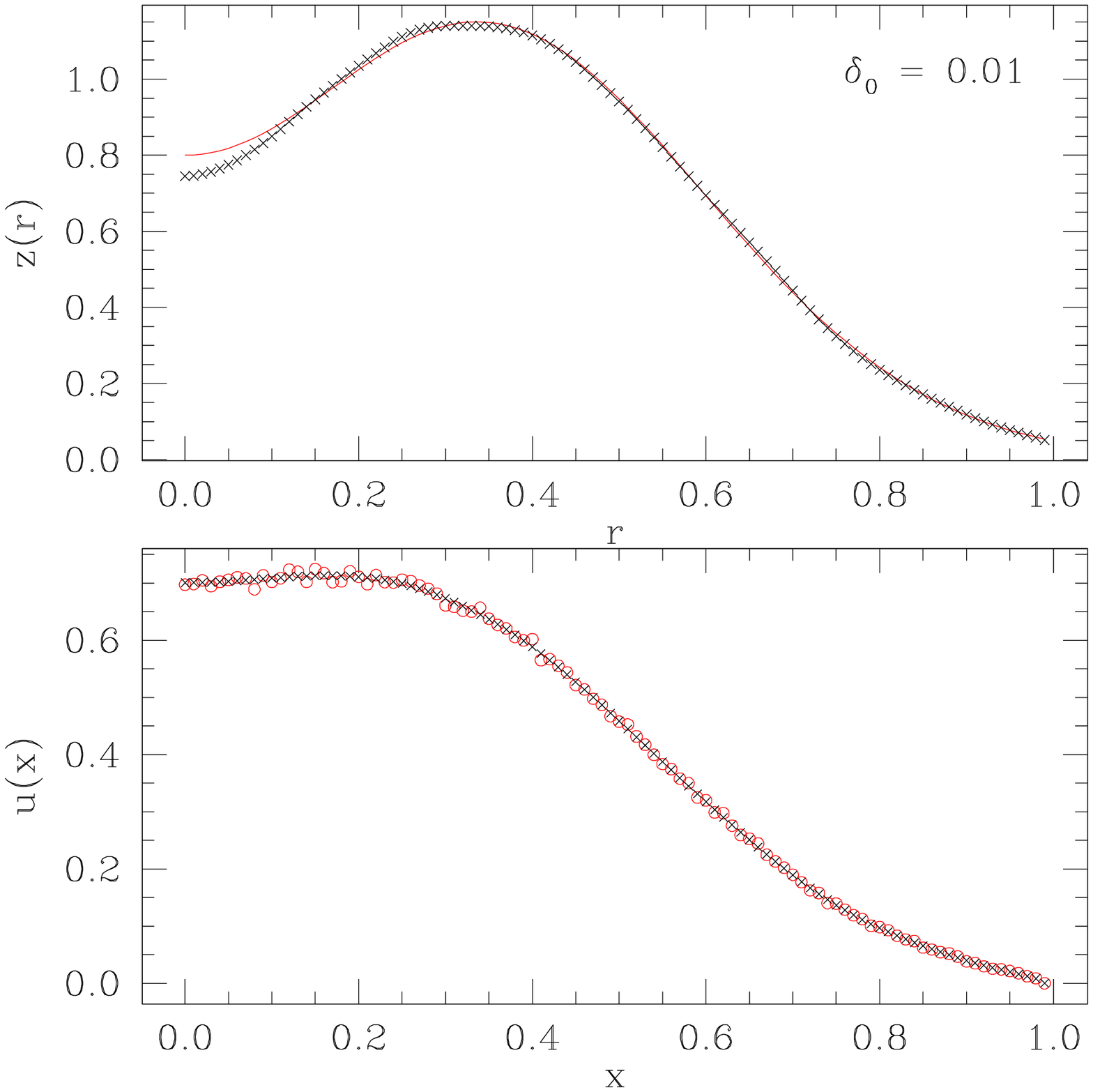}
\caption{Model 3. Tikhonov's regularization without explicit a priori constraints on $u(x)$. Left plot: solution in the metric space $W_2^1$. Right plot: solution in the metric space $W_2^2$.}
\label{fig5}
\end{figure}

Two solutions in the metric spaces $W_2^1$ and $W_2^2$ are shown in Fig.\ref{fig5}. Clearly, the latter solution is more smooth than the former one, for the price of being not so accurate approximation of $z(r)$. This is the result of compromise suggested by Tikhonov's regularization.

\begin{figure*}
\centering
\includegraphics[width=0.3\textwidth]{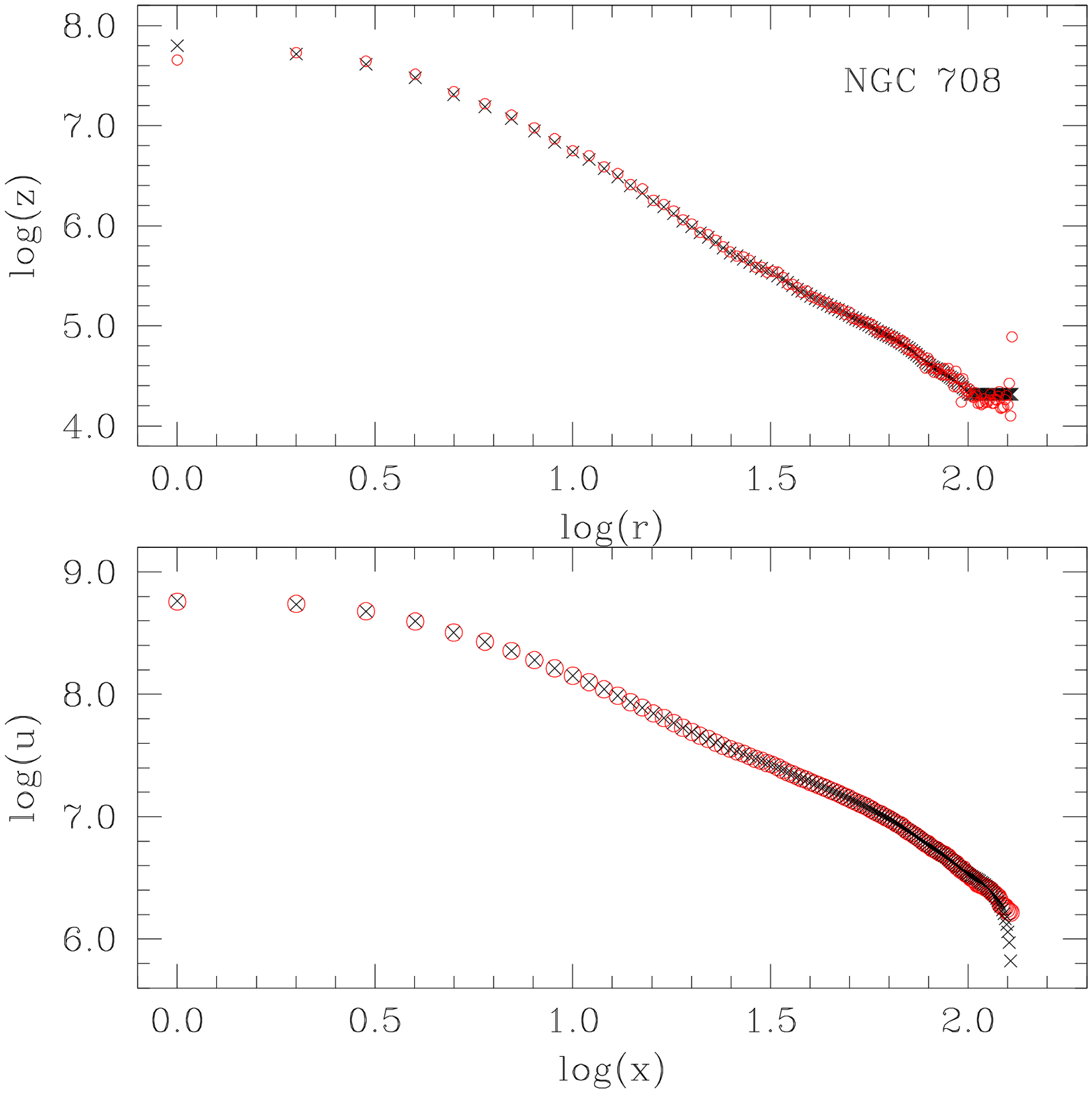}
\includegraphics[width=0.3\textwidth]{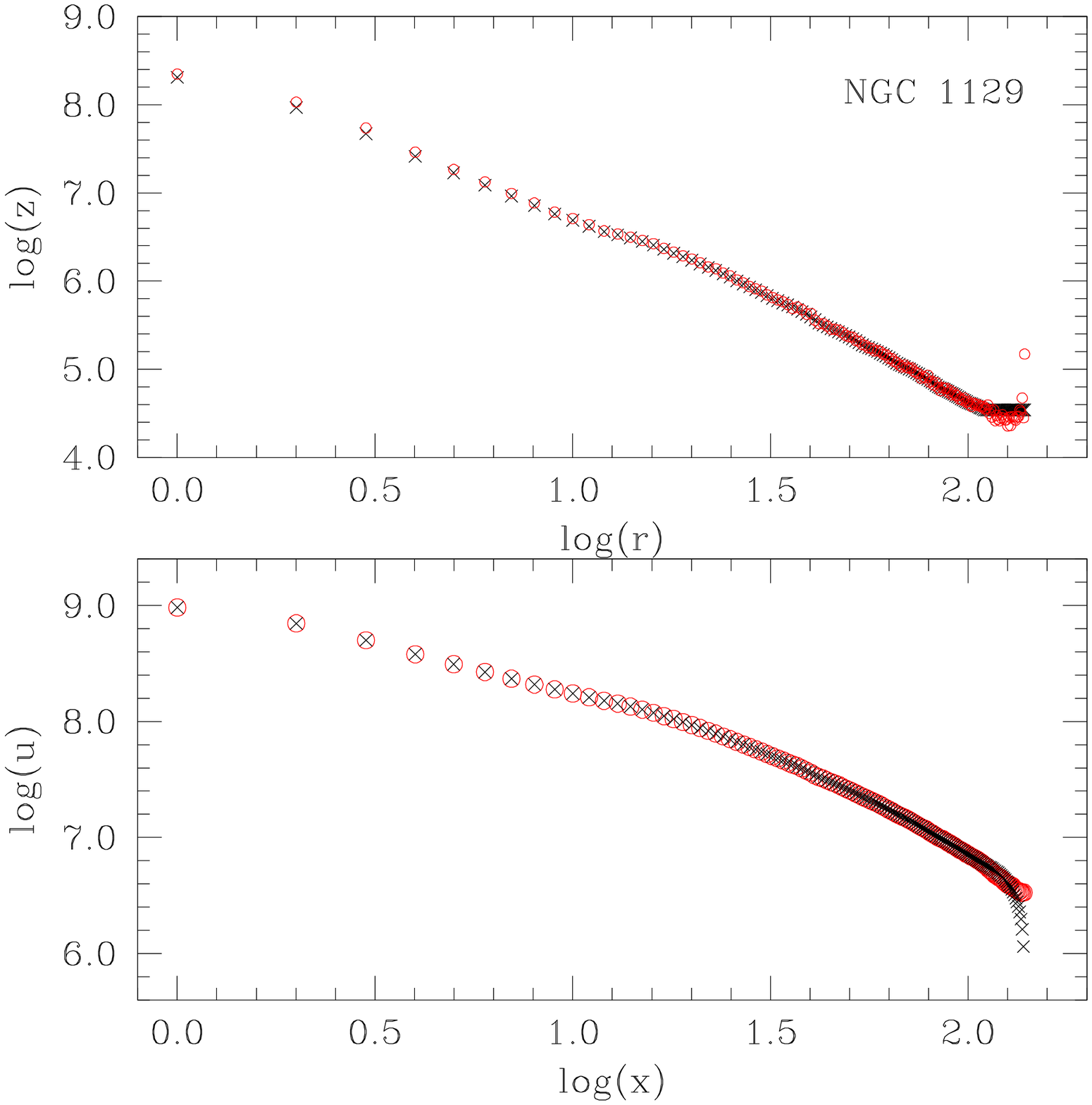}
\includegraphics[width=0.3\textwidth]{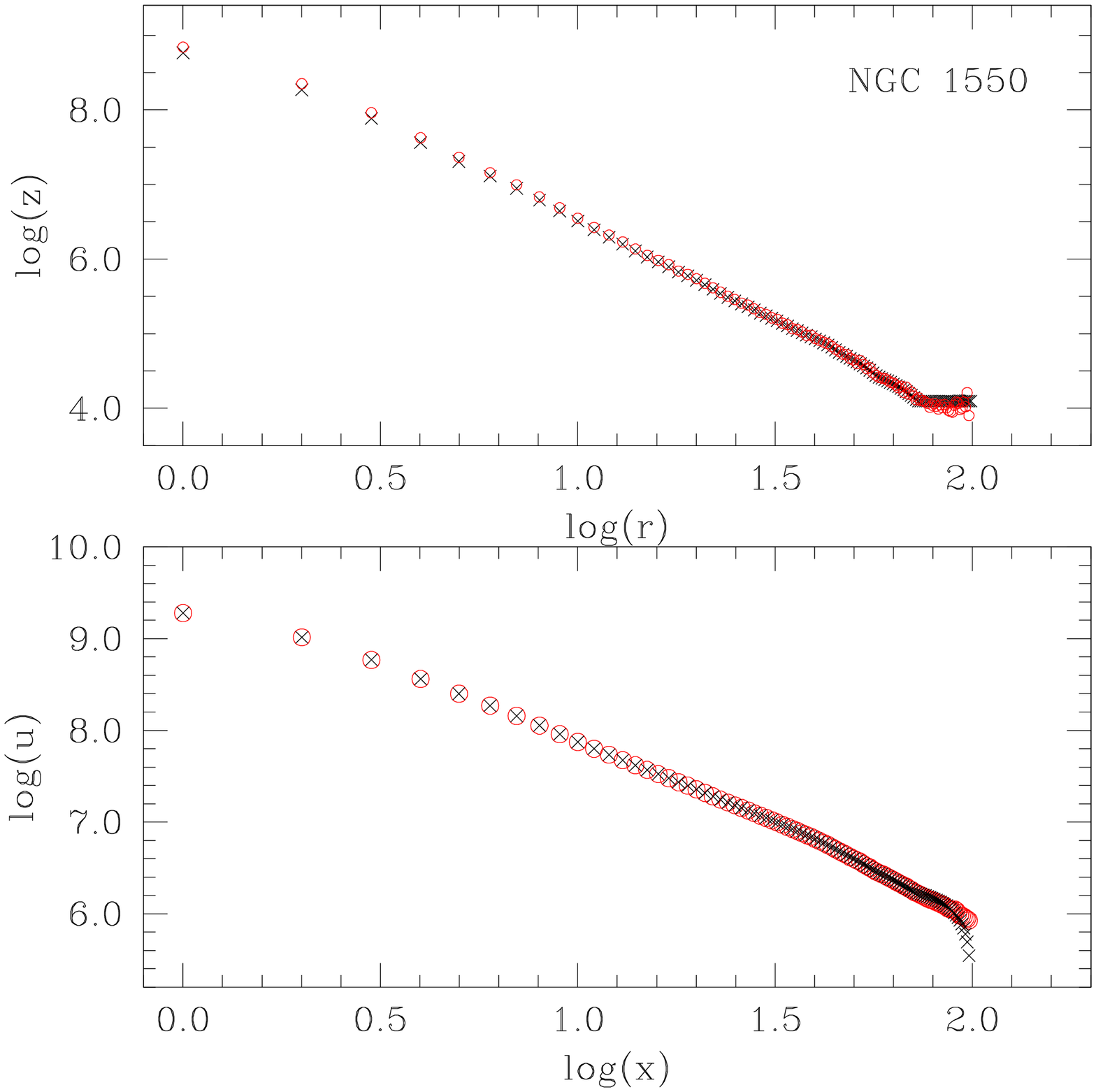}
\caption{Bottom plots: measured (red circles) and model (black crosses) surface density. Upper plots: volume density derived by Wallenquist's method (red circles) and obtained as a result of the current solution (black crosses).}
\label{fig6}
\end{figure*}

\section{An astrophysical example}

\cite{Lyskova2014} discuss kinematics of several early-type elliptical galaxies. The subject is non-trivial as these galaxies are lacking dynamical tracers with known orbits. The authors suggest a method which uses (i)the line-of-sight velocity dispersion and (ii)the surface brightness. The first approach provides dynamic mass, i.e. the total mass including hypothetical dark matter. The second one provides the mass of stellar population. When compared, these values can be used to deduce the fraction of dark matter in a galaxy. The surface brightness can be translated to surface density via the mass-to-light ratio (dependant on stellar population). Knowing the surface density and solving equation \eqref{eq1}, one can obtain the radial distribution of volume mass density. When integrated over the galactic volume, it provides the desired mass estimate for stellar population.

\begin{figure}
\centering
\includegraphics[width=\columnwidth]{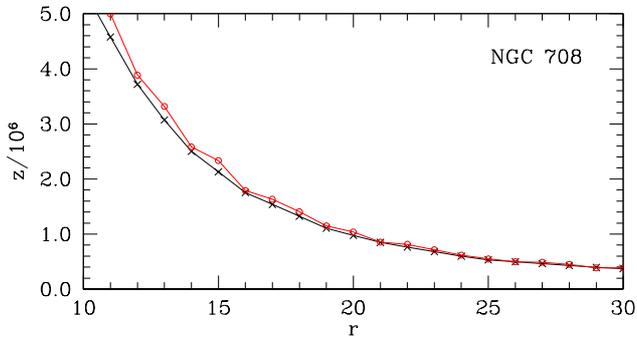}
\caption{A zoomed part (in linear scale) of Fig.\protect\ref{fig6} for NGC~708. \protect\cite{Lyskova2014} solution is plotted as red circles, current solution as black crosses.}
\label{fig7}
\end{figure}

To solve equation \eqref{eq1}, the authors used the \cite{Wallenquist1933} method. In Fig.\ref{fig6}, we compare the solution obtained with our method to the solution by \cite{Lyskova2014}, for three elliptical galaxies.The densities vary by several orders of magnitude, thus a logarithmic scale is used in the plots to emphasize differences between the models. The radii are expressed in arc seconds, the surface density in solar masses per square arc second. Thus, the volume density is in solar masses per volume corresponding to one arc second. In our method, the unknown volume density was assumed to be a monotonically non-increasing and convex (the second derivative is non-negative) function. Tikhonov's regularization was not used. While the overall agreement is good, two differences are evident. The first one is that when using Wallenquist's method, the tail of the volume density function is noisy and non-monotonic, the density increasing with radius. This is non-physical and probably results from large errors in the measured surface density at large radii. In our method, this behaviour is avoided thanks to a priori constraints on the unknown function. A similar problem appears at small radii in Wallenquist's method for NGC~708 (non-monotonic behaviour of the volume density). The second difference is that, in our method, the volume density is indeed a convex function while in Wallenquist's method it is somewhat random, reflecting errors in the input data (see Fig.\ref{fig7}).

In fact, the input data used in this example are of exceptionally high quality. This is why the overall agreement between the current solution and the solution of \cite{Lyskova2014} is so good. As it concerns the stellar galactic mass (volume density integrated over the galactic volume), the difference between the results is practically negligible: the masses obtained with the two methods differ by less than 3 per cent. However, if the detailed shape and smoothness of the unknown function are important, Wallenquist's method may become inadequate.

\section{Conclusions}

We proposed an efficient and flexible method of solving Abel integral equation of the first kind not using any parametrization of the unknown function or smoothing the input data or using more or less arbitrarily chosen approximation of the unknown function by a set of Chebyshev polynomials etc. Instead, the proposed method makes use of a priori constraints on the unknown function based on simple physical considerations and common sense. These constraints can be as weak as to assume that the function is non-negative and monotonically non-increasing (or non-decreasing, then a constant defining the upper limit of the unknown function must be set). If there are grounds to assume more strict constraints like concavity or convexity of the function, these can be used as well. Alternatively, Tikhonov's regularization can be used, which does not require any explicit a priori constraints and instead assumes that the unknown function is smooth. The desired smoothness can be set. Moreover, Tikhonov's regularization can be combined with explicit a priori constraints to obtain a smooth solution satisfying the constraints.

The use of the method is illustrated on a set of simulated models with known solutions, for various functions of interest. It is shown that, in accordance with the theoretical predictions of the method, the approximate solution is approaching the exact one as the errors of the input data decrease. An astrophysical example is also considered illustrating the differences between the current method and that by Wallenquist.

The proposed method can be used in many areas of astrophysics, physics, chemistry, and applied sciences. The computer code (in the C language) for solving Abel integral equation of the first kind is freely available to all interested parties on request.

\section*{Acknowledgements}

The author thanks Prof. O. Sil'chenko for providing original data on surface and volume mass density (obtained by Wallenquist's method) for NGC~708, NGC~1129, and NGC~1550, and the referee Dr. Saha Prasenjit for helpful comments. This research was supported by Russian Foundation for Basic Research through the grant No 14-02-00825 and by Support Program for Science Schools through the grant No  9670-2016.2.

%%%%%%%%%%%%%%%%%%%%%%%%%%%%%%%%%%%%%%%%%%%%%%%%%%

% Don't change these lines
\bsp	% typesetting comment
\label{lastpage}
\end{document}